\RequirePackage{ifpdf}
\ifpdf
\documentclass[pdftex]{sigma}
\else
\documentclass{sigma}
\fi
\usepackage{graphicx}
\usepackage{caption}
\usepackage{subcaption}
\usepackage{mathabx}

\numberwithin{equation}{section}

\newtheorem*{Theorem*}{Theorem}

 { \theoremstyle{definition}

 }

\begin{document}

%\renewcommand{\PaperNumber}{***} %this adds in the paper number on the right of the header

%\FirstPageHeading %this adds in the SIGMA header

\ShortArticleName{A Kac-Weyl Character Identity}
\thispagestyle{empty} %this is what gets rid of the header here
\ArticleName{A Kac-Weyl Character Identity}

% Names of the authors for the title of the paper
\Author{Michael A. BAKER~$^{\rm a}$, Dipesh BHANDARI~$^{\rm b}$and Michael CRESCIMANNO~$^{\rm c}$ }

\AuthorNameForHeading{M. A. Baker, D. Bhandari and M. Crescimanno}

\Address{$^{\rm a)}$~Department of Mathematics, 719 Patterson Office Tower
Lexington, Kentucky 40506-0027, USA} % Address of First Author
\EmailD{\href{mailto:email@address}{mabaker@uky.edu}} 
%\URLaddressD{\url{http://www.home.org/~myHome/}} %URL address of First Author

% Address of Third Author
\Address{$^{\rm b)}$~Department of Physics, Southern Methodist University, Dallas, Texas 75275-0175, USA}
\EmailD{\href{mailto:email@address}{dbhandari@mail.smu.edu}}

% Address of Second Author
\Address{$^{\rm c)}$~Department of Physics and Astronomy, Youngstown State University, Youngstown, OH, 44555, USA}
\EmailD{\href{mailto:email@address}{dcphtn@gmail.com}} 
% In the case of the same organization, please use the following standard
%\Author{First Names LASTNAME and Second COAUTHOR}
%\AuthoqNameForHeading{F.N. Lastname and S. Coauthor}
%\Address{Address of Author(s), Country}
%\Email{\href{mailto:email@address}{email1@address}, \href{mailto:email@address}{email2@address}}
%\URLaddress{\url{URL1}, \url{URL2})

%\ArticleDates{Received ???, in final form ????; Published online ????}

\Abstract{An explicit quantization of Chern-Simons theory leads to an identity between sums of the Kac-Weyl characters. One can use this identity to prove inequalities that constrain the fusion coefficients $N_{\mu\nu}^l$ in the case of RCFTs that descend from current algebras. It also leads to a statement regarding the conjugacy symmetry of the sums of squares of fusion coefficients for current algebras admitting complex representations.}

%Please type here List of Keywords for your article separated by semicolon.

%\Classification{?????; ?????; ?????} % e.g. 35A30; 81Q05
% For 2020 Mathematics Subject Classification see https://mathscinet.ams.org/mathscinet/msc/msc2020.html

\section{Notation} Given a highest weight vector $\mu$, the character $\chi_{\mu}$ is the unsigned exponential sum
over the collection of weights (vectors) $\Omega_\mu$ making up the representation $\mu$. 
\begin{equation} 
\chi_{\mu}({\hat a}_t) = \sum_{r \in \Omega_{\mu}} e^{(r, {\hat a}_t)} = {\rm Tr}_{\mu} (e^{{\hat a}_t}) 
\label{Char}
\end{equation} 
where the $(,)$ is the Cartesian inner product on the weight space and the trace ``Tr" above is with the exponential computed in the representation $\mu$. Here ${\hat a}_t$ is an arbitrary vector in the weight lattice. Throughout, our sums over the weights in $\Omega_\mu$ include the weight's multiplicity. 

Note that each element $w$ of the Weyl group $W$ permutes the weights in the representation, indicating that the above character sums are unchanged under Weyl action on ${{\hat a}_t}$. The Weyl action presages that each character can be realized as a ratio of alternating sums over the Weyl group action on just the highest weight alone (a single orbit of length $|W|$), as  
\begin{equation} 
\chi_\mu({\hat a}_t) = {\frac{D_{\mu+\rho}({\hat a}_t)}{D_{\rho}({\hat a}_t)}} \qquad {\rm where} \qquad
D_{\mu}({\hat a}_t) = \sum_{w\in W} (-1)^w e^{(w(\mu), {\hat a}_t)}
\label{GenChar}
\end{equation} 
where $(-1)^w$ is $\pm 1$, the parity of the $w$ (as an element in the permutation of the simple roots) and the distinguished vector $\rho = {\frac{1}{2}} \sum_{\alpha>0} \alpha$ is also a member of the weight lattice. The $D_{\mu}({\hat a}_t)$ have odd parity under the transpositions that generate the Weyl action on ${{\hat a}_t}$. Importantly for what follows, this explicit form for the character allows one to unambiguously define a notion of `character' for any element of the weight lattice (not just highest weights). It is this generalization we refer to below simply as the character, though in some of the literature they are referred to as `virtual' characters.

\section{Statement of the Theorem} We prove the following: 

{\bf Theorem 2.1. } \textit{For any highest weights} $\mu$ and $\nu$ \textit{we have}
\begin{equation} 
\sum_{\mu' \in \Omega_\mu} \chi_{\mu'+\nu}(k, \tau, u) = \sum_\iota N_{\mu\nu}^\iota \chi_\iota(k,\tau, u)
\label{CBident}
\end{equation} 
in which $\iota$ is a highest weight,  $\chi_{\iota}(k, \tau)$ are the Kac-Weyl characters (for representation $\iota$ at level $k$ and with modular parameter $\tau$) and where $N_{\mu\nu}^\iota$ are the level-$k$ fusion coefficients. Here $u$ is a vector proportional to ${\hat a}_t$.

{\bf Lemma 2.2.} \textit{Taking} $\nu=0$ \textit{(identity representation) leads to the character identity first presented in Ref. \cite{PY2},}
\begin{equation} 
\sum_{\mu' \in \Omega_\mu} \chi_{\mu'} (k,\tau,u) = \chi_\mu (k,\tau,u) 
\label{PYident}
\end{equation} 

The outline for the proof of the theorem is to first display a constructive proof of the following. 

{\bf Lemma 2.3.} \textit{On the variety that supports the chiral ring, the algebra level} $k$ \textit{version of Theorem 2.1 holds: }
\begin{equation} 
\sum_{\mu' \in \Omega_\mu} \chi_{\mu'+\nu}(k,\gamma) = \sum_\iota N_{\mu\nu}^\iota \chi_\iota(k,\gamma) , 
\label{GroupIdent}
\end{equation} 
where $\gamma$ is an element of the variety (note there is no explicit $\tau$ dependence). Once the above lemma is established we show how the  theorem follows by the uniqueness of integration in $\tau$. 

\section{Remarks} 

We provide now a few brief, simple examples and applications to give the reader a more intuitive feel for the theorem and why it is non-trivial (i.e. beyond the abelian case). For simplicity, first take the case $k\rightarrow \infty$ and specialize to the algebra-level limit ($\tau \rightarrow 0, \infty$). For $\mathfrak{su}(2)$ note for the first few representations we have from Eq.~\ref{GenChar} that $\chi_0 = 1$, $\chi_1 = 2\cos(u)$ and $\chi_2 = \cos(2u)+2\cos^2(u)$, $\chi_3 = 4\cos(u)\cos(2u)$, etc. Here, the dimension of the representation $\mu$ is $\mu+1$. By that same definition Eq.~\ref{GenChar}, we have for other, non-highest weights also a rendering of $\chi$ for example, $\chi_{-1}=0$, $\chi_{-2} = -1$, $\chi_{-3} = -2\cos(u)$, that is, for $m>0$, $\chi_{-m} = -\chi_{m-2}$. 

With this abbreviated table, note that, for example, $\sum_{\mu' \in \Omega_1} \chi_{\mu'+1} = \chi_0+\chi_2$ since $\Omega_1 = \{-1,1\}$, and this aligns with fusion on the RHS of Theorem~\ref{CBident} in that $1\otimes 1 = 0 \oplus 2$. Another example: note that $\sum_{\mu' \in \Omega_2} \chi_{\mu'+1} = \sum_{\mu' \in \Omega_1} \chi_{\mu'+2} = \chi_1+\chi_3$ follows from $\Omega_2 = \{-2,0,2\}$ and the fact that $\chi_{-1} = 0$. Note that $\sum_{\mu' \in \Omega_3} \chi_{\mu'+0}$ is a single highest weight term as a result of cancellations due to 
$\chi_{-m} = -\chi_{m-2}$. 

Next we use the fact that there is a positive definite Hermitian norm in the space of characters that makes them an orthonormal basis. Since the multiplicities $m_{\mu'}$ on $\mu$ can be defined as $\sum_{\mu'\in\Omega_{\mu}} m_{\mu'} = \text{dim}(\mu)$ and $\sum_{\mu'\in\Omega_{\sigma}} m_{\mu'} = \text{dim}(\sigma)$, where $m_{\mu'}\in\mathbb{Z}^+$, taking the norm-square of both sides of Eq.~\ref{CBident}, gives 
%us, $\left(\sum_l N^l_{\sigma\mu}\chi_l\right)^2 = ]\left(\sum_{\mu'\in\Omega_{\mu}}\chi_{\sigma+\mu'}\right)^2  \leq \sum_{\mu'\in\Omega_{\mu}} m^2_{\mu'}$. So $\forall k$, 
\begin{align}\sum_l \left( N^l_{\sigma\mu}\right)^2 \leq \min\left(\sum_{\mu'\in\Omega_{\mu}} m_{\mu'}^2, \sum_{\mu'\in\Omega_{\sigma}} m_{\mu'}^2\right) 
\label{ineq}
\end{align} 
true $\forall k$ level. 

One may see that Eq.~\ref{CBident}  is like a generalized Fourier decomposition and  that Eq.~\ref{ineq} is the associated Parseval's identity. Note for example for SU$(3),$ $\sum_l \left( N^l_{3\sigma}\right)^2 \leq 3, \sum_l \left( N^l_{8\sigma}\right)^2 \leq 10$ and $\sum_l \left( N^l_{6\sigma}\right)^2 \leq 6$.
 
Denote the list of representations that make up the fusion ring by $\Omega_{all}$ and form the distinguished vector $v = \sum_{\mu\in \Omega_{all} } \chi_\mu$. Now in the space of characters calculate the inner product of $v$ with  both sides of Eq.~\ref{CBident}. This leads to $\min(\text{dim}(\nu),\text{dim}(\mu))\geq \sum_l N_{\mu\nu}^l$. This inequality allows us to have a constraint that is based on current algebras as the $\text{dim}(\nu)$, etc. is from Eq.~\ref{CBident} the Lie algebra ($k\rightarrow \infty$) dimension and not the quantum dimension of the representation. As such, we do not have a useful generalization of Eq.~\ref{CBident} to RCFTs that may not descend from current algebras (cf. Ref. \cite{dovgard} and Ref. \cite{pinto}). 

Next, let $k$ be finite but keep $\tau \rightarrow 0, \infty$. For $\mathfrak{su}(2)_k$ the characters are $\chi_n = \sin(\pi (n+1) u/(k+2))/\sin(\pi u/(k+2))$. Now the theorem's RHS is evaluated on the $\mathfrak{su}(2)_k$ variety and the level-$k$ fusion $N_{\mu\nu}^l$ emerges. For one simple example, let $k=2$ and consider $\sum_{\mu' \in \Omega_2} \chi_{\mu'+2} = \chi_0+\chi_2+\chi_4 = \chi_0$ where the last inequality follows at $k=2$ from the fact that the variety is the collection of the 8th roots of unity on which $\chi_4 = -\chi_2$. 

Another straightforward application of the theorem is in the conjugacy symmetry of the sums of fusion coefficients. Limit now our discussion to current algebras admitting complex representations (For Lie algebras, $A_n$, $D_n$ and $E_6$). In Ref.~\cite{zuber}, the authors show that $\sum_l N_{ab}^l = \sum_l N_{a{\bar b}}^l$ for any $a, b$ ($\bar b$ denotes the conjugate representation to $b$). Since these sums are just the result of the product of $a$ and $b$ written in terms of characters paired with $v = \sum_{\mu\in \Omega_{all} } \chi_\mu$ through the inner product, the equality $\sum_l N_{ab}^l = \sum_l N_{a{\bar b}}^l$ combined with Eq.~\ref{CBident} automatically indicates that, whether with $b$ or ${\bar b}$, the LHS has the same number of non-zero terms. Thus, appealing to the associated Parseval's identity we recover in this case the result $\sum_l (N_{ab}^l)^2 = \sum_l (N_{a{\bar b}}^l)^2$ without directly appealing to crossing symmetry as in Ref. \cite{zuber}. 

Since all the denominators in Eq.~\ref{CBident} are the same, focusing just on the numerators, we see that the alternating sum over the Weyl group then can be interchanged with the sum over the weight vectors in the representation. Although $\Omega_\mu$ is Weyl invariant, the alternating sum for $D_{\mu}({\hat a}_t)$ is over the weights displaced by $\rho$. Thus, in the resulting sum, only displaced weights that are part of an orbit of length $|W|$ will lead to a non-zero contribution. The difficulty is that in general there will be many of these which apparently according to Eq.~\ref{CBident} actually cancel.

Rather than tracing through the intricacy of these many cancellations, below we show that in the group theory context,  Eq.~\ref{CBident} can be understood via the quantization of Chern-Simons (CS) theory as a diagonalization of the fusion ring of $G_k$. We show that relating these to the (genus 1) modular primaries of the associated WZW model then leads to a proof of the proposed identity for the Kac-Weyl characters. 

\section{Brief Review of Chern-Simons Quantization on $T^2 \times \mathbb{R}$ }
For context, note that the lemma arose as a generalization of understanding the detailed direct quantization of CS theory \cite{PY1}. We adopt an earlier explicit quantization formulation of CS theory \cite{crescimannoHotes} on the $T^2 \times \mathbb{R}$ that was useful for making connections between varieties, chiral potentials and fusion in $G_k$ and coset models \cite{crescimannoGH}, and to a universal formula for the inverse of the handle operator in $G_k$ \cite{crescimannoK}, the quadratic form in which the representations are orthonormal that was alluded to before. One way to view this CS quantization approach is as a explicit implementation of the Racah–Speiser algorithm using operators on a finite Hilbert space.

Briefly, for compact gauge groups, a gauge transformation of the CS action can lead to terms proportional to the gauge group volume. As a quantum theory the integrand of the path integral should not change under this transformation, leading to the conclusion that the overall coefficient of the action must be $k/{2\pi}$ with $k\in {\mathbb Z}$ called the level ($\hbar = 1$). In the canonical quantization framework, the Hamiltonian reduction of the CS theory on $T^2 \times \mathbb{R}$ proceeds via requiring the gauge fields to satisfy the first class constraint $F=0$, that is, the gauge fields are flat. On the torus this indicates that classically the remaining gauge covariant degrees of freedom may be taken to be in the Cartan subalgebra and spatially constant. To quantize the theory it is then necessary to choose a polarization; a simple choice for the 2-torus being along its the principal axes. Let $A_\mu$ ($\mu = 1,~2)$ be the Cartan subalgebra valued gauge field on the 2-torus. The CS action being first order in derivatives then implies for its quantum mechanical description that 
\begin{equation} 
[A^m_1(x), A^j_2(x)] = \frac{2\pi i}{k}  (C^{-1})_{mj} \delta^{(2)}(x-y)
\label{PB1} 
\end{equation}
where $m,j \in {1,2, \ldots, {\rm rank}(G)}$ and $C$ is the Cartan matrix of $G$. The quantization of the theory then consists of constructing a Hilbert space that supports a faithful linear representation of these gauge covariant operators. We call this Hilbert space the Gaussian model, since it, very roughly, can be thought of as that of a tensor products of free fields with constraints. Then gauge invariant operators can be amalgamated from the gauge covariant operators. 

An expectation of CS topological field theory is that there exists a unique ``vacuum" state $\psi_0$ and a ($1-1$) operator-state correspondence, that is, $\psi_{j} = {\cal O}_j \psi_0$ for $\{\psi_j\}$ that span the Hilbert space. In CS quantization as described above (in terms of non-gauge invariant operators) the Hilbert space of the CS theory will be an invariant subspace of the Gaussian Hilbert subspace described above.  

Quite generally, gauge invariant operators are identified as Wilson loop operators in the CS theory, explicitly ${\cal O}_j = {\rm Tr}_j\left(e^{i\int_c A}\right)$ where $j$ labels the representation of the Lie algebra in which the trace is taken and `$c$' labels a closed path. In the associated Gaussian model we instead work directly with the non-gauge invariant components of the Wilson loop operators, $a_j = e^{i\int A^j_1{\rm d}x}$
and $b_m = e^{i\int A^m_2{\rm d}y}$, $x$ and $y$ being a Cartesian co-ordinate along the principal directions of the spatial torus (and each integral being along that entire homotopy loop). 

As linear operators on the Gaussian Hilbert space, the $a_j$  can be multiplied by each other, forming a closed commutative ring called the Gaussian fusion ring; the same is true of the $b_j$ products themselves. The product of the $a_j$ and $b_m$ operators however must support a representation of the commutator Eq.~\ref{PB1}
\begin{equation} 
a_mb_ja_m^{-1}b_j^{-1} = e^{{\frac{2\pi i (C^{-1})_{mj}}{(k+c)}}} 
\label{ClockQuant} 
\end{equation} 
where, for reasons that will become clear later, we have shifted the level $k$ by the Casimir element of the adjoint representation. Since Eq.~\ref{ClockQuant} implies that all the operators are idempotent, a finite minimal Gaussian Hilbert space $\{ |\gamma\rangle \}$ to support a faithful linear representations of these unitary operators can be built up by taking $a_j$ as diagonal and $b_m$ as `shift' operators. The Gaussian vacuum state  $ |0\rangle $ is defined via $a_j|0\rangle =|0\rangle$ for all $a_j$ and $\langle 0| 0\rangle = 1$. The action of the operators $a_j$ and $b_m$ in the Gaussian model is $b_j|v\rangle = |v+{\hat e}_j\rangle$ and 
$a_j|v\rangle = \exp\left({\frac{2\pi i (C^{-1} {\vec v})_j}{k+c}}\right)|v\rangle$
We denote the Hilbert space of the Gaussian model by $\Lambda_{k+c}$ or for brevity suppress the subscript and just denote it $\Lambda$ with the understanding that it depends on the level. 

The Wilson loop operators (gauge invariant) are then particular polynomials in the $a_j$ or  $b_m$. The states in associated rational conformal theory $\psi_r = {\frac{1}{\sqrt{|W|}}}\sum_{w\in W} (-1)^w \Pi_j b_j^{w(r+\rho)_j} |0\rangle = {\cal O}_r({\bf b}) |\psi_0\rangle$ with $|\psi_0\rangle$ the unique vacuum state $\psi_0 = {\frac{1}{\sqrt{|W|}}}\sum_{w\in W} (-1)^w \Pi_j b_j^{w(\rho)_j} |0\rangle$ in the CS/conformal correspondence and displaying the expected state-operator correspondence in the RCFT.  These formulae indicate the RCFT Hilbert space is identified as the fully Weyl-odd subspace of the Gaussian model whereas the Wilson loop operators  ${\cal O}_r$ are Weyl even. Explicitly  ${\cal O}_\mu(b) = \sum_{v\in \Omega_\mu} \Pi_{j} b_j^{v_j}$. One may think of Weyl transformation as  `large' gauge transformations; here the operators are then gauge invariant whereas the states are gauge covariant and parity-odd under Weyl's primitive permutations, a choice that still always leads to gauge invariant expectation values. 

Fusion in $G$ ($k\rightarrow \infty$) is via the commutative ring ${\cal O}_r {\cal O}_s = N_{rs}^t(\infty) {\cal O}_t$. On the variety Eq.~\ref{ClockQuant}  however, by virtue of the associated idempotency of the $b_j, a_j$, this commutative ring truncates to that of $G_k$, namely ${\cal O}_r({\bf b}) {\cal O}_s({\bf b}) = N_{rs}^t {\cal O}_t({\bf b})$ where $N_{rs}^t$ are the level $k$ fusion coefficients (note the same as $N_{rs}^t(\infty)$). The $a_j$ and $b_k$ were themselves related to the gauge degrees of freedom by a choice of polarization, a different (physically equivalent) choice---for example due to a modular transformation---will be related to this choice by a unitary transformation of the Gaussian Hilbert space which will then faithfully restrict to a unitary transformation on the RCFT Hilbert space. For example for the  $S$ transformation, ($a_j\rightarrow b_j$ and $b_j\rightarrow a_j^{-1}$); since the $a_j$ all commute with one another, they are the diagonal representation of the $b_j$'s, that is, $S^{-1}b_jS = a_j$. 

The Hilbert space of the RCFT has a natural norm it inherits from the underlying `Gaussian model,' that is, $\langle0|0\rangle=1$ and the operator algebra on the unitary operators $b_i$ , $a_j$ implies that $\langle\psi_r|\psi_s\rangle = \delta_{rs}$. 

\section{Proof of Lemma 2.3} A constructive proof of the lemma 2.3 then proceeds via realizing the characters as a mixed inner product; that is, between a state in the Gaussian Hilbert space and one in its RCFT Hilbert (sub-)space. Choose $|\gamma\rangle \in\Lambda$ as any state the Gaussian model.  Form
\begin{equation} 
\langle\gamma|S^{-1}|\psi_\mu\rangle = \langle \gamma|S^{-1} {\cal O}_\mu(b)|\psi_0\rangle = \langle\gamma|{\cal O}_\mu(a) S^{-1}|\psi_0\rangle 
\label{starter}
\end{equation} 

where note that ${\cal O}_\mu(b)$ acts on the weight space through a sum of translations (by the weights in $\Omega_\mu$).  Given $\psi_0 = {\frac{1}{\sqrt{|W|}}}\sum_{w\in W} (-1)^w \Pi_j b_j^{w(\rho)_j} |0\rangle$, then $S^{-1}|\psi_0\rangle = {\frac{1}{\sqrt{|W|}}}\sum_{w\in W} (-1)^w \Pi_j a_j^{w(\rho)_j} S^{-1}|0\rangle$ and, on general grounds as described earlier, in the Gaussian model note $S^{-1}|0\rangle =  {\frac{1}{\sqrt{|\Lambda_{k+c}|}}} \sum_{{\vec l\in \Lambda}}  |{\vec l}\rangle$, a finite sum over the entire level-$k$ sublattice in the weight space defining the Gaussian model. We arrive at a rendering of the character as an inner product in the Hilbert space, 
\begin{align} 
\langle\gamma|S^{-1}|\psi_\mu\rangle  &= {\frac{1}{\sqrt{|\Lambda_{k+c}|}}} {\frac{1}{\sqrt{|W|}}}\sum_{w\in W} (-1)^w \sum_{\vec l} \langle\gamma|\Pi_j a_j^{w(\rho + \mu)_j} |l\rangle\\ 
&=  {\frac{1}{\sqrt{|\Lambda_{k+c}|}}} {\frac{1}{\sqrt{|W|}}}\sum_{w\in W} (-1)^w \exp\left({\frac{2\pi i ({\vec \gamma}C^{-1} ({\vec {\mu}+{\vec \rho}}))}{k+c}}\right)\\ &= {\frac{1}{\sqrt{|\Lambda_{k+c}|}}} {\frac{1}{\sqrt{|W|}}}
D_{\mu+\rho}(\gamma) 
\label{AlmostFinal}
\end{align} 
Next,  using this for ${\cal O}_\nu({\bf b}) {\cal O}_\mu({\bf b}) = \sum_\iota N_{\nu \mu}^\iota {\cal O}_\iota({\bf b})$ we have
\begin{equation}
\langle\gamma|S^{-1}{\cal O}_\mu({\bf b})|\psi_\nu\rangle = {\frac{1}{\sqrt{|\Lambda_{k+c}|}}} {\frac{1}{\sqrt{|W|}}}\sum_\iota N_{\nu \mu}^\iota
 D_{\iota+\rho}(\gamma) 
\label{fusionKet}
\end{equation} 
Likewise using representation fusion to write
\begin{align} 
\langle \gamma|S^{-1} {\cal O}_\mu(b)|\psi_\nu\rangle &= \sum_{\mu'\in \Omega_\mu} \langle\gamma| \Pi_j a^{\mu'_j}S^{-1}|\psi_\nu\rangle\\&= {\frac{1}{\sqrt{|W|}}} \sum_{\mu'\in \Omega_\mu} \sum_{w\in W} (-1)^w \langle\gamma|\Pi_j a^{\mu'_j+w(\nu+\rho)_j} S^{-1}|0\rangle\\ 
 &=  {\frac{1}{\sqrt{|\Lambda_{k+c}|}}} {\frac{1}{\sqrt{|W|}}}\sum_{\mu'\in \Omega_\mu} \sum_{w\in W} (-1)^w \exp\left({\frac{2\pi i {\vec \gamma}C^{-1} ({\vec \mu'}+{\vec \nu}+{\vec \rho})}{k+c}}\right)
\label{ConstrFinal}
\end{align} 
where for the last equality we used the fact that the $\Omega_{\mu}$ is even under $W$ and necessarily consists of disjoint $W$--orbits (perhaps of different lengths but) in which each element has the same multiplicity. 
Clearly combining Eq.~\ref{fusionKet} and Eq.~\ref{ConstrFinal} leads to the theorem (sans common denominator) for the $G_k$ in that, 
\begin{equation} 
\sum_{\mu'\in \Omega_\mu} D_{\mu'+\nu+\rho} (\gamma)= \sum_\iota N_{\nu \mu}^\iota
 D_{\iota+\rho}(\gamma) 
\label{GroupFinal}
\end{equation}
Recall $\gamma$ was an arbitrary state in the Gaussian model, so Eq.~\ref{GroupFinal} is the lemma 2.3 evaluated on the variety associated with the intersection of polynomials Eq.~\ref{ClockQuant}. 

%\begin{figure}[!ht] %changed from htbp to !ht
%\centerline{\includegraphics[width=6.3in, height=2.85in, trim={0.05in 2.38in 0.05in 0.001in},clip]{figure1.eps}}
%\caption{This is an example of a figure}
%\label{fig1}
%\end{figure}

\section{Proof of the Theorem: The Kac-Weyl Characters} 

The lemma 2.3 can be extended to an identity on the associated Kac-Weyl characters. We follow the methods and notation of \cite{elitzur}. The starting point is to look at a multiplet of doubly periodic functions on the torus with modular parameter $\tau$, so that $x \rightarrow x+1$ and $x\rightarrow x+\tau$ delineate the spatial symmetries we mod the plane out to arrive at the torus. Clearly, to be everywhere finite and avoid strict periodicity on the torus (which would leave only the constant section), we generalize to projective periodicity. A useful starting point in that regard is to 
limit ourselves to projective representations of the forgoing spatial symmetry. 

To relate this covariance to the functions of the topological field theory, we can realize it as a linear redefinition of the gauge fields. Thus, in the topological field theory, the wavefunctions are valued in a space that has the rank of the gauge algebra. Let $u$ be an arbitrary vector in that space and let $\alpha, \beta$ represent root vectors there. A projective representation that leads to a modular representation can, without loss of generality, be written in terms of functions $f(\tau, u)$ satisfying
\begin{equation} 
f_k(\tau, u+\beta) = f_k(\tau, u) \qquad \qquad f_k(\tau, u+\tau\beta) = e^{-i\pi k \tau (\beta,\beta) -2\pi i (\beta, u)}f_k(\tau,u)
\label{germ1}
\end{equation} 
where $k$ is a natural number. Let $\gamma$ represent a weight vector. For the ADE Lie algebras, it is straightforward to show that the `Gaussian' sum,
\begin{equation}
\Theta_{\gamma, k}(\tau, u) = \sum_{\alpha\in \Lambda^R} e^{i\pi k\tau (\alpha+\frac{\gamma}{k})^2 + 2\pi i k (\alpha+\frac{\gamma}{k}, u) }
\label{theta}
\end{equation}
satisfies Eq.~\ref{germ1}, with $u$ replacing $\gamma$ by linear superposition of fourier components. Here $\Lambda^R$ is the integer lattice generated by the positive roots, and by $(\alpha+\frac{\gamma}{k})^2$ we mean the length squared of the vector. For this $\Theta_{\gamma, k}(\tau, u)$, the first equation of Eq.~\ref{germ1} follows from the (usual) root normalization $(\alpha_i, \alpha_j)=\frac{(\alpha_i,\alpha_i)}{2} C_{ij}$ and the second follows from the fact that $\Lambda^R$ is preserved by shifts by any root. 

So defined, these $\Theta_{\gamma, k}(\tau, u)$ satisfy  
\begin{equation} 
\Theta_{\gamma, k}(\tau+1, u) = e^{i\pi (\gamma, \gamma)/k} \Theta_{\gamma, k}(\tau, u) \qquad 
\Theta_{\gamma, k}(-1/\tau, u) = \frac{ e^{-i\pi k(u+\frac{\gamma}{k})^2}}{\xi {\rm det}(-kC/\tau)^{\frac{1}{2}}} {\tilde \Theta}_{\gamma, k}(\tau, \tau u)
\label{thetaModular}
\end{equation}
where the former uses the integrality of $(\alpha,\gamma)$ and the later is a lattice generalization of the Poisson resummation formula so that ${\tilde \Theta} (\tau, \tau u)$ is defined as in Eq.~\ref{theta} but where the sum there is over the scaled weight lattice $\Lambda^w/k$,  $\xi$ is the primitive eighth root of unity (\cite{mumford}, Eq. 5.6 pg. 195) and $C$ the Cartan matrix. 

With $\Theta_{\gamma, k}(\tau, u)$ so defined, it is now straightforward to construct---in terms of them---the Kac-Weyl characters. We do that in terms of symmetric and antisymmetric sums over the Weyl group's action. Define
\begin{equation} 
\Theta_{\gamma, k}^+(\tau, u)  = \sum_{w\in W} \Theta_{w(\gamma), k}(\tau, u) \qquad \qquad \Theta_{\gamma, k}^-(\tau, u)  = \sum_{w\in W} (-1)^w \Theta_{w(\gamma), k}(\tau, u)
\label{weylTheta1}
\end{equation}
The Kac-Weyl characters $\chi$ are then
\begin{equation} 
\chi_{\gamma, k}(\tau, u) = \Theta_{\gamma+\rho, k+c}^-(\tau, u)/\Theta_{\rho, k+c}^-(\tau, u) 
\label{weylTheta2}
\end{equation} 
And we can now readily relate these characters to $D$ of the Lie algebra approach in the earlier section. First note that the $\Theta$ satisfy the parabolic second order differential equation, 
\begin{equation} 
\biggl(\nabla_u^2 -4\pi ik \frac{\partial}{\partial \tau} \biggr) \Theta_{\gamma, k}(\tau, u) = 0
\label{parabolic} 
\end{equation} 
which by linearity is true for $\Theta^-_{\gamma, k}(\tau, u)$ as well and as written develops no singularities when evolving from initial data at, say, $\tau=0$. Now forming up the identity Eq. \ref{CBident}, multiplying both sides by the common denominator $\Theta^-_{\rho, k+c}(\tau, u)$ we can then use the linearity and  uniqueness of the solution of the Eq. \ref{parabolic} to compare the two sides of the identity. They will be the same if their boundary (initial data) agree. We then note that in the $\tau \rightarrow 0 $ limit that $\Theta^-_{\mu+\rho, k+c}(\tau, u) \rightarrow D_{\mu+\rho} (u)$ where $u$ here is the analytic extension of the evaluation of the $D_\mu (\gamma)$ from the values $\gamma$ in the variety defined via Eq. \ref{ClockQuant}.

Now, to further one's intuition, we perform an explicit check of identity \ref{CBident}  for $\Theta^-_{\gamma, k}(\tau, u)$ in the case of $\mathfrak{su}(2)_2$ to further elucidate the argument we made about the Kac-Weyl extension and we do so without resorting to arguments about uniqueness and boundary conditions. The generalised characters for the $\mathfrak{su}(2)_k$ have numerators that are: 
\begin{align}
    \chi_j \sim \Theta^-_{j+1, k+2}(\tau,u) = \sum_{\alpha\in\mathbb{Z}} e^{2i\pi \tau\left(\alpha+\frac{j+1}{k+2}\right)^2 + 4\pi i  \left(\alpha+\frac{j+1}{k+2}\right)u} - e^{2i\pi\tau\left(\alpha-\frac{j+1}{k+2}\right)^2 + 4\pi i  \left(\alpha-\frac{j+1}{k+2}\right)u}, 
\label{gench} 
\end{align}
where $j = 0, \dots, k$ labels representations of  the $k+1$ conformal blocks. By a shift in $\alpha$ note also that for representation labels $j$ and $m$ if $j+m>k+1$ the $\chi_{j+m} = -\chi_{2(k+1)-j-m}$ (implying that $\chi_{k+1}=0$). 

Without loss of generality take $j > m$. If $j+m<k+1$, we expect from the LHS of theorem's statement that  $j\otimes m = j-m\oplus j-m+1\oplus \ldots \oplus j+m$, if however $j+m>k+1$ then $j\otimes m = j-m\oplus j-m+1 \oplus \ldots \oplus 2k-j-m$. These are the expected $\mathfrak{su}(2)_k$ fusion ring relations.

\section{Conclusion}
Take a representation $\sigma$. A generalization of the character formula to non-highest weight vectors allows us to write the sum over the weight vectors in a representation $\mu$ of the generalized characters at $\sigma$ displaced by those weight vectors as equal to a sum over highest weight characters in the tensor product, complete with multiplicities. This can be understood via the rather explicit canonical quantization of the associate Chern-Simons (CS) theory defined on $T^2 \times {\mathbb R}$. 

These considerations lead to a version of the identity for Kac-Weyl characters, as well as a bound on the sums of squares of the  fusion coefficients in terms of the  dimensions of the participating representations. One avenue for future exploration is whether there exists a canonical way to generalize the characters of fusion algebras that are not related to current algebras/CS theory so that the analogous  Eq.~\ref{CBident} still holds. It would be likewise worth investigating other bounds on the fusion coefficients and reaching a deeper understanding of their utility and generality as that may lead to insight in the counting/classification of RCFTs \cite{bruillard}. 

%\appendix

%\section{First appendix}
%We kindly ask the authors to give all technical details of paper in the form of Appendices.

\subsection*{Acknowledgements}

The authors acknowledge partial support via NSF grant DMR-2226956. We acknowledge helpful discussions with M. Porrati. 

%\bibliographystyle{sigma}
%\bibliography{example}

\pdfbookmark[1]{References}{ref}
\LastPageEnding

\end{document}